\documentclass[aps,prb,twocolumn,showpacs]{revtex4}

\usepackage{amsmath}
\usepackage{amsfonts}
\usepackage{amssymb}
\usepackage{graphicx}
\usepackage{dsfont}

\newcommand{\abs}[1]{\ensuremath{\left\vert#1\right\vert}}

\newcommand{\heaviside}[1]{\ensuremath{\theta\left[#1\right]}}
\renewcommand{\Re}[1]{\ensuremath{\operatorname{Re} \left\{#1\right\} } }
\renewcommand{\Im}[1]{\ensuremath{\operatorname{Im} \left\{#1\right\} } }

\renewcommand{\vec}[1]{\boldsymbol{#1} }

\begin{document}

\title{Interplay between spin-orbit interactions and
a time-dependent electromagnetic field in monolayer graphene}
\author{Andreas Scholz}
\email[To whom correspondence should be addressed. Electronic
 address:]{andreas.scholz@physik.uni-regensburg.de}
\author{Alexander L\'{o}pez}
\author{John Schliemann}
\affiliation{Institute for Theoretical Physics, University of Regensburg, D-93040 Regensburg, Germany}
\date{\today}

\begin{abstract}
We apply a circularly and linearly polarized terahertz field on a monolayer of graphene
taking into account spin-orbit interactions of the intrinsic and Rashba types.
It turns out that the field can be used not only to induce a gap in the energy spectrum,
but also to close an existing gap due to the different reaction of the spin components
with circularly polarized light.
Signatures of spin-orbit coupling in the density of states of the driven system can be observed
even for energies where the static density of states is independent of spin-orbit interactions.
Furthermore it is shown that the time evolution of the spin polarization and the orbital dynamics
of an initial wave packet can be modulated by varying the ratio of the spin-orbit coupling parameters.
Assuming that the system acquires a quasi stationary state,
the optical conductivity of the irradiated sample is calculated.
Our results confirm the multi step nature of the conductivity obtained recently,
where the number of intermediate steps can be changed by adjusting the 
spin-orbit coupling parameters and the orientation of the field.
\end{abstract}


\maketitle

\section{Introduction}
Since a monolayer of graphite was isolated and detected for the first time,\cite{Novoselov_2004}
many theoretical and experimental studies on this remarkable and surprising material
have been published.\cite{Neto_2009}
Even several years after its discovery graphene remains one of the most intense
research topics in solid state physics.
This expresses the high expectations and hopes physicists have for graphene 
as being a building block for novel electronic devices.

While in the beginning the focus of graphene research was mainly set on spin-independent phenomena,
as it was claimed that spin-orbit interactions (SOIs) are virtually unimportant in graphene,\cite{Kane_2005, Gmitra_2009}
recent experimental and theoretical works have demonstrated that spin-orbit coupling (SOC) effects might 
be important as the characteristic parameters can be enlarged significantly.
\cite{Neto09, Weeks_2011, Ma_2012, Dedkov_2008, Varykhalov_2008, Jin_2012}
This, in principle, opens up the possibility of using spin-related phenomena in this 
outstanding material with exceptional electronic properties.
Moreover, and indeed fortunately, graphene is not the only promising two-dimensional hexagonal system and thus many of the 
findings of spin-related research in graphene can also be applied to other systems.
As an example we mention a monolayer of MoS$_2$\cite{Mak_2010, Radis_2011, Zeng_2012} which can 
at low energies effectively be described as two uncoupled gapped graphene systems,
where both the band gap and the SOIs turn out to be large,
\cite{Xiao_2012, Li_2012, Kormanyos_2013}
and silicene, a two-dimensional honeycomb lattice made of silicon atoms.\cite{Liu_2011, Chen_2012, Vogt_2012}
Furthermore, restricting ourselves to nearest-neighbor and interlayer hopping,
the Hamiltonian of bilayer graphene is formally equivalent to that of monolayer graphene
with a purely Rashba SOC,
where the Rashba coefficient is substituted by the interlayer hopping constant.

In recent works the effects of an external time-dependent field 
on two-dimensional materials such as a monolayer
\cite{Calvo_2011, Oka_2009, Oka_2011, Zhou_2011, Busl_2012, Lopez_2012, Calvo_2012, Delplace_2013}
or bilayer\cite{Morell_2012} of graphene, 
HgTe/CdTe quantum wells,\cite{Lindner_2011, Cayssol_2012}
or \textit{n}- and \textit{p}-doped electron gases\cite{Fujita_2003, Cheng_2005, Jiang_2008, Zhou_2008}
have been discussed.
It was shown that an electromagnetic field can induce
gaps in the energy spectrum of graphene\cite{Calvo_2011, Oka_2009, Oka_2011, Zhou_2011}
and even move and merge the Dirac points.\cite{Delplace_2013}
Both aspects might be interesting for future applications such as transistors.
Furthermore, in Ref.~\cite{Lindner_2011} the possibility of changing the topology of a 
HgTe/CdTe quantum well by applying linearly polarized light leading to so-called Floquet
topological insulators has been reported.

In this work we study how SOIs of the intrinsic and Rashba types
manifest themselves in a monolayer of graphene under the influence
of a time-dependent field whose energy is on the terahertz (THz) regime.
The SOC coupling constants are chosen to be of the order of the photon energy.
This work is organized as follows.
In Sec.~\ref{Sec:Model}, we introduce the model Hamiltonian and briefly summarize the main
properties of the solution of Schr\"odinger's equation according to Floquet's theorem.
In Sec.~\ref{Sec:Spectrum}, the energy spectrum and the density of states (DOS)
of the driven system is discussed
and signatures arising from the interplay of SOC and the THz field are pointed out.
In Sec.~\ref{Sec:Dynamics}, the dynamics of physical observables such as the spin polarization
and the position operators are studied.
In Sec.~\ref{Sec:Conductivity}, the optical conductivity of the irradiated sample
is calculated for various combinations of the SOC parameters.
Finally, Sec.~\ref{Sec:Conclusions} summarizes the main results of this work.

\section{The model}
\label{Sec:Model}
We use the Kane-Mele model\cite{Kane_2005} (setting $\hbar = 1$ throughout this work)
\begin{align}
\hat H_0 = v_F\vec k \cdot \vec\sigma s_0 + \lambda_I \sigma_z s_z + \lambda_R \left( \vec\sigma \times \vec s \right) \vec e_z
\label{Hamiltonian_static}
\end{align}
to describe a monolayer of graphene including SOIs of the intrinsic ($\lambda_I$) and Rashba ($\lambda_R$) types at one $K$ point.
The Pauli matrices $\vec\sigma$ ($\vec s$) and the unit matrix $\sigma_0$ ($s_0$) act on the pseudospin (real spin) space.
The other $K$ point can be described by the above Hamiltonian with $\sigma_x\rightarrow-\sigma_x$ and $\sigma_z\rightarrow-\sigma_z$.
The effect of the electromagnetic field can be incorporated by the minimal coupling scheme $\vec k \to \vec k + e\vec A(t)$.
As the vector potential does not depend on the position operators, the Hamiltonian remains diagonal in
momentum space and we can treat $\vec k$ as a number instead of a differential operator.
The time-dependent contribution to the Hamiltonian,
\begin{align}
\hat H_1(t) = ev_F\vec A(t) \cdot \vec \sigma s_0 ,
\label{Hamiltonian_driving}
\end{align}
is assumed to be periodic in time, i.e., $\hat H_1(t+T) = \hat H_1(t)$, where $T = 2\pi / \Omega$
and $\Omega$ is the frequency of the radiation field.

The vector potential describing a monochromatic wave propagating perpendicular to the graphene plane
can be assumed to be either classical,
\begin{align}
\vec A(t) = \frac {\sqrt{2} E_0}\Omega \left[ \cos{\theta_p} \cos{\Omega t} ~\vec e_x 
+ \sin{\theta_p} \sin{\Omega t} ~ \vec e_y \right] ,
\end{align}
or quantized,
\begin{align}
\vec A(t) &= \mathcal A \left[ \cos{\theta_p} \left( \hat a e^{-i\Omega t} + \hat a^\dagger e^{i\Omega t} \right) ~ \vec e_x 
\right. \notag \\
&\left. \qquad + i \sin{\theta_p} \left( \hat a e^{-i\Omega t} - \hat a^\dagger e^{i\Omega t} \right) ~ \vec e_y \right] ,
\end{align}
where, among obvious notation, the parameter $\mathcal A$ contains geometric information about the cavity
surrounding the system.
The field is either circularly ($\theta_p = 45^\circ$) or linearly polarized 
(e.g., along the $x$ direction for $\theta_p = 0^\circ$).
Quantizing the electromagnetic field adds a degree of freedom described by the bosonic operators $\hat a^{(\dagger)}$,
which comes along with a new conserved quantity given by the helicity
$\hat h = \hat J + \hat a^\dagger \hat a $,
where the angular momentum
$\hat J = x k_y - y k_x + \sigma_z / 2$
generates rotations of the carrier degrees of freedom in real and pseudospin space.
To treat the electromagnetic field as a quantized operator is important in situations
where the charge carriers have a significant back-action on the field which in turn can alter
the particle dynamics itself.

To analyze this aspect further, let us consider the case of a vanishing field and neglect SOIs for the moment.
Now assuming a wave packet with initial momentum along the $y$ axis
and the pseudospin initially in the $x$ direction,
the dynamics of the system in the Heisenberg representation is given by
\cite{Schliemann_2005, Schliemann_2006, Schliemann_2007, Schliemann_2008, David_2006, Trauzettel_2007, David_2010}
\begin{align}
\frac {d^2}{dt^2} x_H(t) 
= -2 v_F^2 k \sin{(2v_Fkt)} 
\end{align}
and $d^2 y_H(t) / dt^2 = 0$.
From the classical expression for the radiative power of dipolar radiation,\cite{Landau_Fields}
$P = (e \ddot{\vec r})^2 / 6 \pi \epsilon_0 c^3$,
we find the time-averaged energy loss per time as
\begin{align}
\bar P = \frac {e^2 v_F^4 k^2} {3\pi \epsilon_0 c^3}
\approx 7.12 \times 10^{-2} \, \frac {k^2} {\text{nm}^{-2}} \frac {\text{meV}}{\text{ps}} ,
\label{Radiative_power}
\end{align}
i.e., for a wave vector of $k = 0.1$ nm$^{-1}$ the radiative power is of order 
$10^{-4}$ meV/ps.
Due to the very large Fermi velocity of $v_F = 10^6$ m/s in graphene, a time scale of 1 ps corresponds,
for appropriate initial conditions, to a distance of 1 $\mu$m traveled by the wave packet.
Therefore, the above loss rate should be seen as a small effect.\cite{Note_Quantization_in_2DEG}
Hence the energy loss due to dipolar radiation induced by \textit{Zitterbewegung}
can be neglected compared to other energy scales in typical experimental situations.
Accordingly, in what follows we will treat the electromagnetic field as a classical quantity and not
as an operator.
For convenience we introduce the dimensionless quantity $\alpha = v_F e E_0 / \Omega^2$.

Due to the periodicity of $\hat H(t) = \hat H_0 + \hat H_1(t)$, 
the solution of Schr\"odinger's equation
$\left[ i \partial/\partial_t  - \hat H \right] \left\vert \Psi_{\vec k, \mu\nu} \right\rangle = 0$
obeys Floquet's theorem\cite{Grifoni_1998, Kohler_2005} and thus is of the form
\begin{align}
\left\vert \Psi_{\vec k, \mu\nu} \right\rangle = 
e^{-i\varepsilon_{\vec k, \mu\nu} t} \left\vert \psi_{\vec k, \mu\nu} (t) \right\rangle ,
\label{wavefunction_Floquet}
\end{align}
where $\mu, \nu = \pm1$ are band indices.
The Floquet states $\left\vert \psi_{\vec k, \mu\nu} (t) \right\rangle$ have the same periodicity
as the Hamiltonian and can be expanded in a Fourier series:\cite{Shirley_1965}
\begin{align}
\left\vert \psi_{\vec k, \mu\nu} (t) \right\rangle = \sum_{n=-\infty}^\infty e^{in\Omega t} 
\left\vert \chi_{\vec k,\mu\nu}^n \right\rangle .
\end{align}
The original problem can now be reduced to the diagonalization of the time-independent
Floquet Hamiltonian whose components are defined by
\begin{align}
\left(\hat H_{F}\right)_{nm} = \frac 1T \int_0^T dt \, \hat H(t) e^{i(n-m)\Omega t} - n\Omega \delta_{nm} .
\label{Floquet_Hamiltonian}
\end{align}
The time evolution of an arbitrary state with respect to an initial time $t_0$
is captured by the operator
\begin{align}
\hat U_{\vec k} (t, t_0) = \sum_{\mu',\nu'} e^{-i\varepsilon_{\vec k,\mu'\nu'}(t-t_0)}
\left\vert \psi_{\vec k, \mu'\nu'} (t) \right\rangle \left\langle \psi_{\vec k, \mu'\nu'} (t_0) \right\vert .
\label{TE_operator}
\end{align}
Notice that the energies and wave functions entering Eq.~(\ref{wavefunction_Floquet}) are not uniquely defined as
$\left\vert \Psi^n_{\vec k, \mu\nu} \right\rangle = e^{in\Omega t} \left\vert \Psi_{\vec k, \mu\nu} \right\rangle$
(with $n\in \mathbb{Z}$) is a solution of Schr\"odinger's equation as well.
The corresponding quasienergy $\varepsilon^n_{\vec k,\mu\nu} = \varepsilon_{\vec k,\mu\nu} + n\Omega$
differs only by a multiple of the THz energy.
Hence the choice of the eigenenergies is ambiguous as they describe the same physical situation.
In order to get a well-defined quantity that is the same for all $\varepsilon^n_{\vec k,\mu\nu}$,
we furthermore introduce the time-averaged (or quasi stationary) energy\cite{Hsu_2006, Faisal_1996, Gupta_2003, Zhou_2011}
\begin{align}
\bar \varepsilon_{\vec k, \mu\nu} = \frac 1T \int_0^{T} dt  
\left\langle \Psi_{\vec k, \mu\nu}(t) \vert \hat H(t) \vert \Psi_{\vec k, \mu\nu} (t) \right\rangle .
\label{Mean_energy}
\end{align}
In general, there is a non trivial relation between the quasienergies and the mean energies.
Notice that in the absence of the driving Eq.~(\ref{Mean_energy}) reproduces the energies
of the unperturbed system (see below).

\begin{figure}[t]
\includegraphics[scale=0.32]{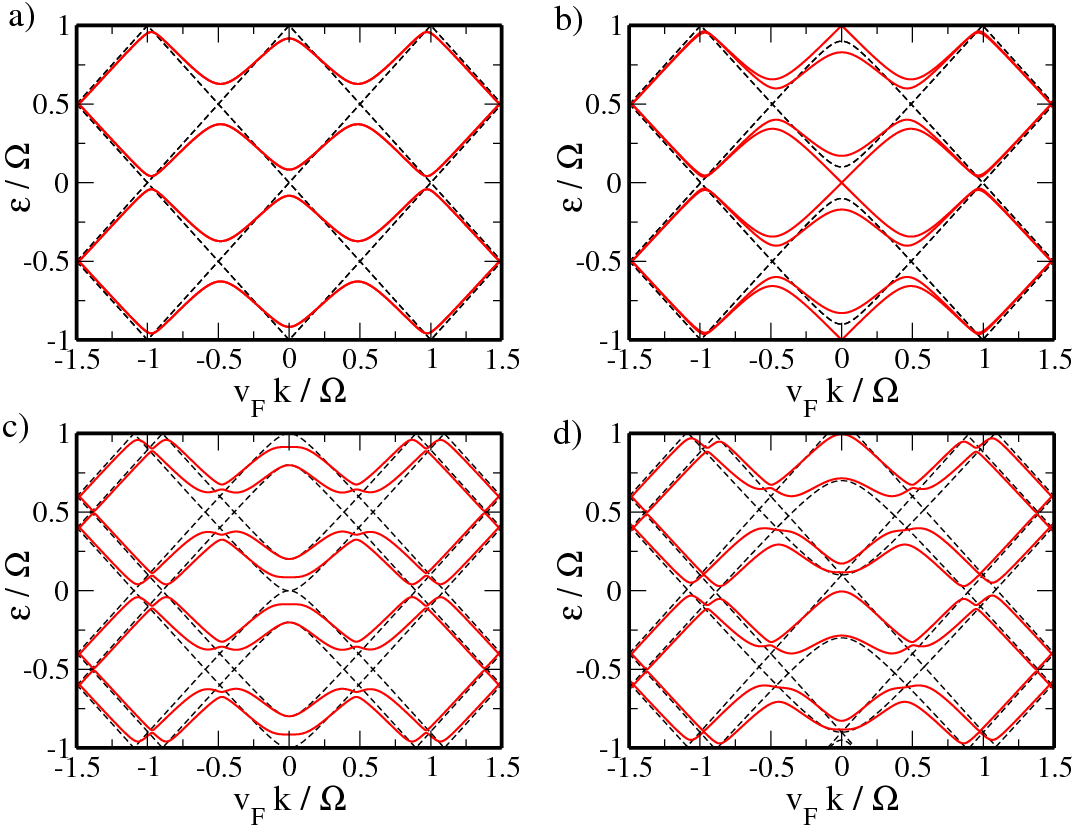}
\caption{(Color online) Quasienergy spectrum under circularly polarized light ($\theta_p = 45^\circ$)
for various combinations of the SOC parameters:
$(\lambda_R/\Omega, \lambda_I/\Omega) = $ (a) $(0, 0)$, (b) $(0, 0.1)$, (c) $(0.1, 0)$ (c), and (d) $(0.1, 0.1)$.
The field strength was set to $\alpha = 0.3$.
}
\label{FIG_spectrum}
\end{figure}

\section{Energy spectrum and density of states}
\label{Sec:Spectrum}
\subsection{Energy bands}
The energy bands of the static problem ($\alpha = 0$) can readily be obtained:
\begin{align}
E_{\mu\nu} (k) = \mu \; \lambda_R + \nu \; \sqrt{v_F^2k^2 +  \left(\lambda_R - \mu \lambda_I\right)^2 } .
\label{Static_energies}
\end{align}
For a finite driving, the eigensystem is calculated numerically by diagonalization of the 
Floquet Hamiltonian in Eq.~(\ref{Floquet_Hamiltonian}).
As mentioned above, this leads to an infinite number of eigenenergies\cite{Shirley_1965} where 
only four of them are physically independent (corresponding to the dimension of the problem),
while all others can be obtained by adding or subtracting a
multiple of the energy of the electromagnetic field $E_{em} = \Omega$.

\begin{figure*}[t]
\includegraphics[scale=0.3]{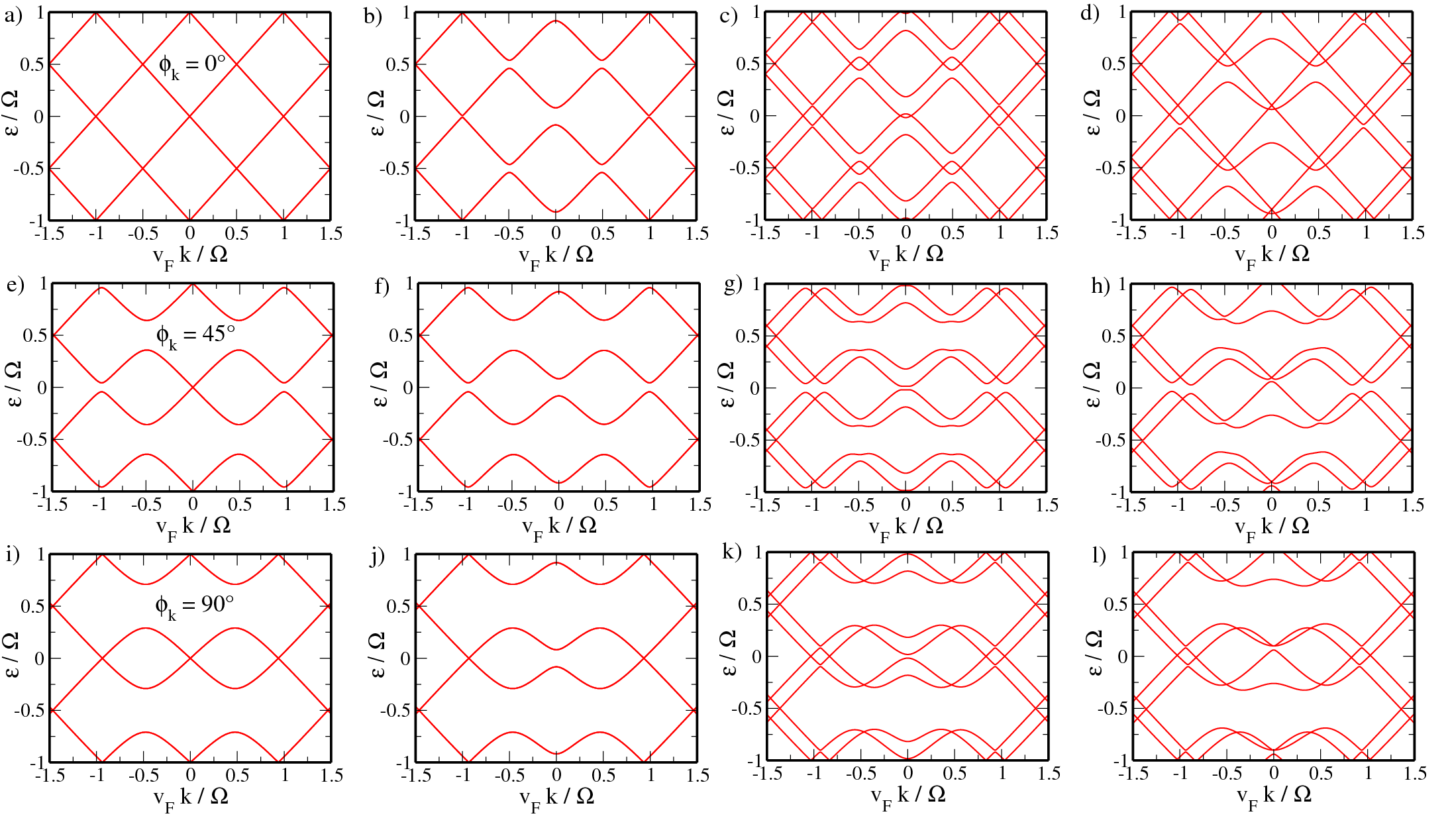}
\caption{(Color online) Quasienergy spectrum under linearly polarized light ($\theta_p = 0^\circ$)
for various combinations of the SOC parameters
and momentum in plane orientations ($\tan \phi_k = k_y / k_x$):
$(\lambda_R/\Omega, \lambda_I/\Omega) = $ $(0, 0)$ (left column), $(0, 0.1)$ (second from left),
$(0.1, 0)$ (second from right), and $(0.1, 0.1)$ (right).
The field strength was set to $\alpha = 0.3$.
}
\label{FIG_spectrum_lin_pol}
\end{figure*}

In Figs.~\ref{FIG_spectrum} and \ref{FIG_spectrum_lin_pol} the quasienergies within the first and second Brillouin zones (BZs)
are shown as red lines for different combinations of the SOC parameters for a fixed field strength of $\alpha = 0.3$.
The black dashed line in Fig.~\ref{FIG_spectrum} shows, for comparison, Eq.~(\ref{Static_energies})
projected to the BZ.
The SOC parameters are chosen to be of the order of the THz energy, e.g., in the present case $\lambda_{R/I} \sim 0.1\Omega$.
As mentioned in the introduction, it has been demonstrated that $\lambda_R$ and $\lambda_I$ can be enlarged 
by several orders of magnitude by choosing proper adatoms\cite{Neto09,Weeks_2011, Ma_2012}
or a suitable environment\cite{Dedkov_2008, Varykhalov_2008, Jin_2012} which allows values of $\lambda_R$
and $\lambda_I$ in the THz (meV) range.
Our results depend only on the ratio $\lambda_{R/I}/\Omega$ and on the coupling strength $\alpha$.
Hence they may also be applied to fields with larger frequencies (such as the mid-infrared)
provided the SOC parameters are large enough.
The advantage of a THz field, however, is that the field energies are far below the energies 
of optical phonons (of about $200$ meV),\cite{Ando_2007} such that excitations of optical phonons are suppressed.

\begin{figure}[b]
\includegraphics[scale=0.32]{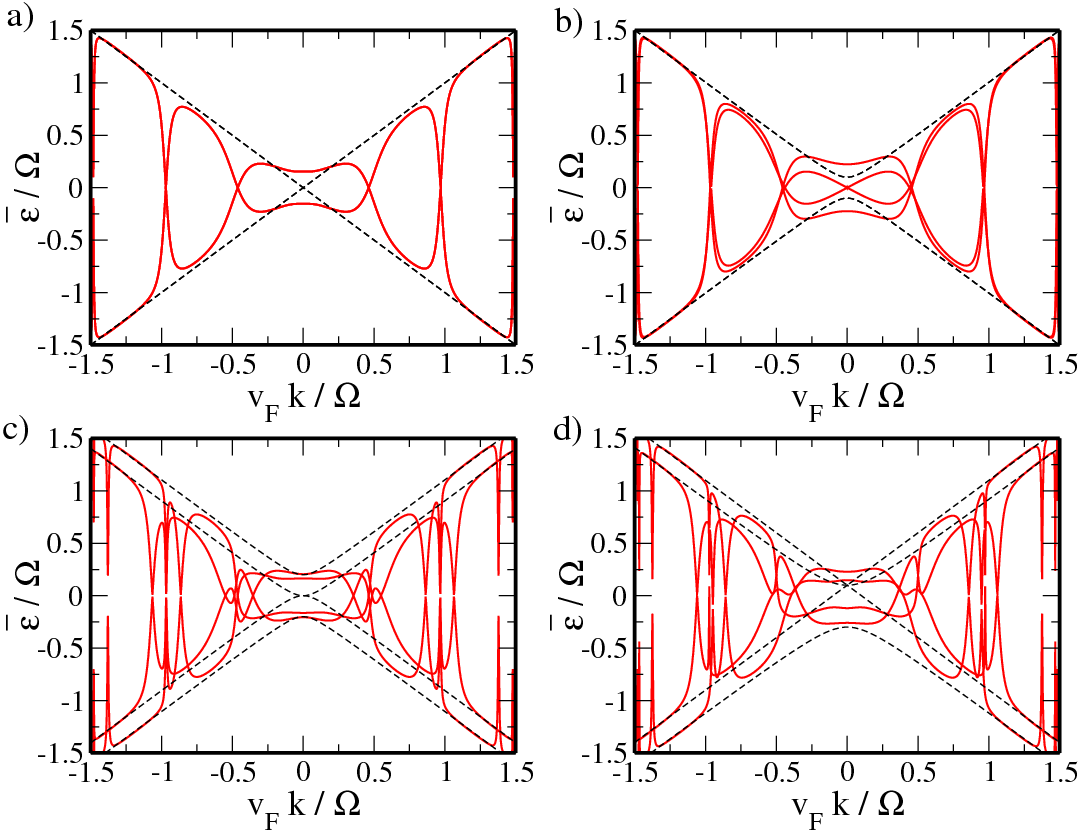}
\caption{(Color online) Mean energies derived from Eq.~(\ref{Mean_energy})
under circularly polarized light ($\theta_p = 45^\circ$) for various combinations of the SOC parameters:
$(\lambda_R/\Omega, \lambda_I/\Omega) = $ (a) $(0, 0)$, (b) $(0, 0.1)$, (c) $(0.1, 0)$, and (d) $(0.1, 0.1)$.
The field strength was set to $\alpha = 0.3$.
}
\label{FIG_average_energy}
\end{figure}

%%%%%%%%%%%%%%%%%%%
%%%%%%%%%%%%%%%%%%%
\textit{Circular polarization.}\quad
The unperturbed energy spectrum of Eq.~(\ref{Static_energies}) consists of twofold spin-degenerate bands
if $\lambda_R = 0$, while for a finite Rashba coefficient structure inversion symmetry is broken
and the bands split up; see the dashed lines in Fig.~\ref{FIG_spectrum}.
Once $\hat H_1(t)$ is turned on, in Figs.~\ref{FIG_spectrum}(a),
\ref{FIG_spectrum}(c) and \ref{FIG_spectrum}(d) a gap opens up right the Dirac point,
separating the valence and conduction bands.
Here the bands are parabolic around the \textit{K} point
but closely follow the linear behavior of the unperturbed result for $v_F k \gtrsim \Omega$.
Similarly, a finite gap also appears in the mean energies lifting the \textit{K} point degeneracy,
e.g., in Fig.~\ref{FIG_average_energy}(a) with a gap of $\bar\delta_0 = 4\Omega\alpha^2 / \sqrt{1 + 4\alpha^2}$.
For finite SOIs the bands react differently on the THz field
and hence the degeneracy present in the static case of Fig.~\ref{FIG_spectrum}(b),
where $\lambda_I = 0.1\Omega$ and $\lambda_R = 0$, disappears.
Right at the Dirac point the quasienergy gap vanishes,
while a new gap opens up between the conduction (or valence) band states with different spin orientations.
Two of the four bands are now linear and not parabolic as in the case of $\alpha = 0$.
Similarly, the gap in the time-averaged energies shown in Fig.~\ref{FIG_average_energy}(b)
is closed.
For larger momenta, $v_F k > 0.7 \Omega$, the spin splitting in Fig.~\ref{FIG_average_energy}(b) eventually becomes so small that
the bands are virtually degenerate again.

\begin{figure*}[t]
\includegraphics[scale=0.3]{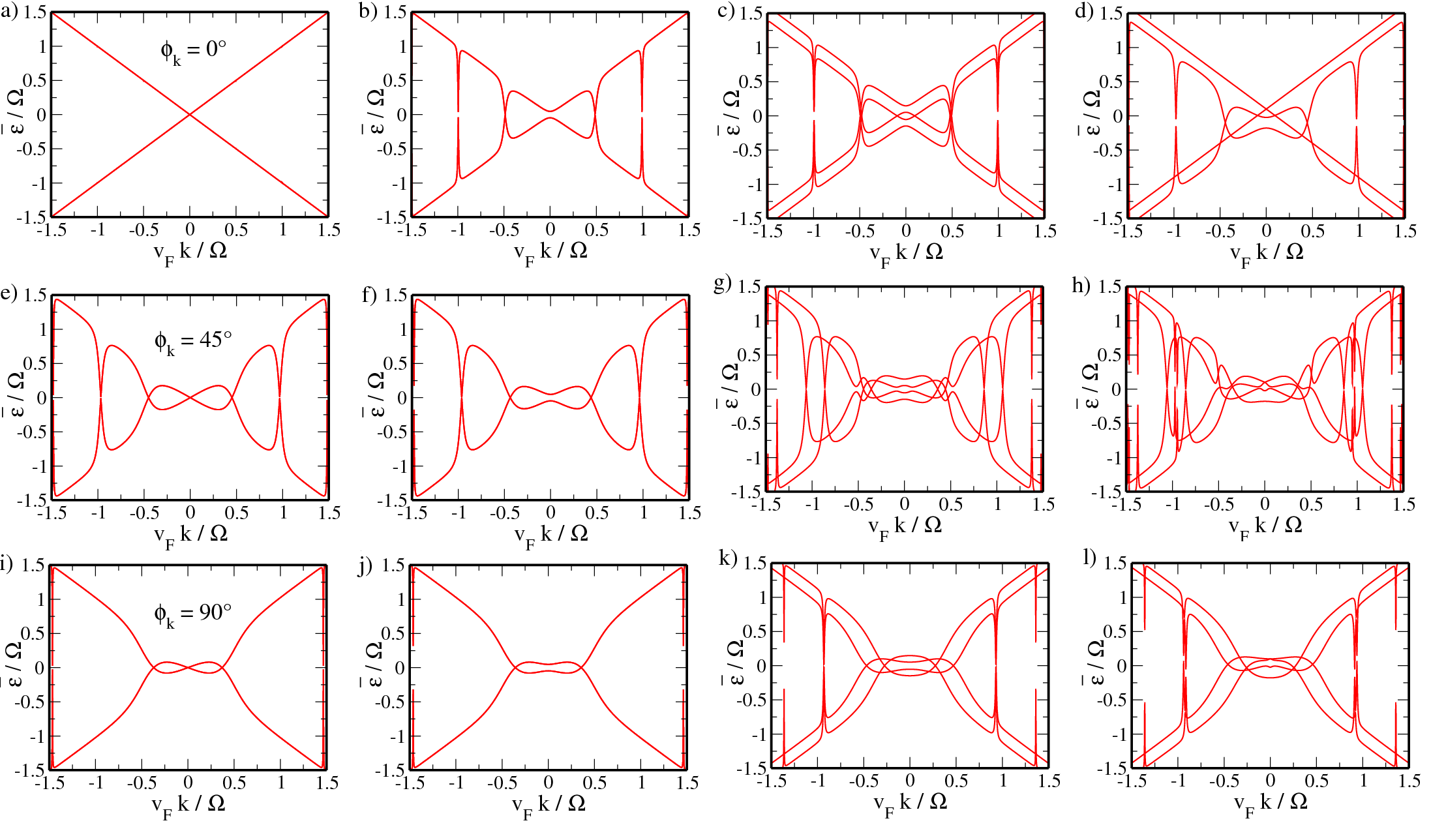}
\caption{(Color online) Mean energies derived from Eq.~(\ref{Mean_energy}) under 
linearly polarized light ($\theta_p = 0^\circ$) for various combinations of the SOC parameters
and momentum in plane orientations ($\tan \phi_k = k_y / k_x$):
$(\lambda_R/\Omega, \lambda_I/\Omega) = $ $(0, 0)$ (left column), $(0, 0.1)$ (second from left),
$(0.1, 0)$ (second from right), and $(0.1, 0.1)$ (right).
The field strength was set to $\alpha = 0.3$.
}
\label{FIG_avg_energy_lin_pol}
\end{figure*}

From Fig.~\ref{FIG_spectrum} we can see that besides the gap at the Dirac point,
additional gaps appear at $v_F k \approx n \Omega / 2$ ($n\in \mathbb{Z}$).
While these gaps are quite large for $v_Fk \approx \Omega / 2$ and $\Omega$, its value
strongly decreases for larger momenta and seems to vanish for $v_F k \gtrsim 1.5\Omega$.
The reason for these gaps is the existence of photon resonances,\cite{Zhou_2011}
i.e., the absorption and emission of photons,
similar to the ac Stark effect in semiconductors.\cite{Grifoni_1998}
Here transitions might occur at the resonant points
$E_{\mu\nu} - E_{\mu'\nu'} \approx n E_{em}$.
In the vicinity of the resonances, $v_F k \approx 0.5n\Omega$, the average energies drop to zero.
For large enough momenta the dips eventually become so narrow that they seem to disappear.
In case spin degeneracy is broken (i.e., $E_{+\pm} \neq E_{-\pm}$),
the above resonant condition can be fulfilled for multiple values of $k$
and hence we observe not one but several nearby dips in the average energy spectrum,
as shown in Figs.~\ref{FIG_average_energy}(c) and \ref{FIG_average_energy}(d).

\textit{Linear polarization.}\quad
If the field is linearly polarized, in the following along the $x$ direction,
the energy spectrum is expected to be strongly anisotropic.
In contrast to the circular case spin degeneracy is broken only if $\lambda_R \neq 0$.

From Fig.~\ref{FIG_spectrum_lin_pol}(a) we can see that for $\lambda_{R} = 0$ and $\lambda_I = 0$
the quasienergy spectrum for an in-plane angle of $\phi_k = 0^\circ$,
where $\tan\phi_k = k_y / k_x$, exactly follows the unperturbed spectrum,
i.e., the field has no influence.\cite{Zhou_2011}
However, if at least one of the SOC parameters is finite,
the valence and conduction bands no longer touch at $v_F k \approx 0.5\Omega$,
where deviations from the static results are largest,
and the THz field induces a gap as shown in Fig.~\ref{FIG_spectrum_lin_pol}(b).
The corresponding time-averaged energies, shown in Figs.~\ref{FIG_avg_energy_lin_pol}(b)-\ref{FIG_avg_energy_lin_pol}(d),
exhibit characteristic dips at $v_F k \approx 0.5n\Omega$,
as for circularly polarized light.
However, from Figs.~\ref{FIG_avg_energy_lin_pol}(a) and \ref{FIG_avg_energy_lin_pol}(d)
we can see that only those bands are affected by the THz field
that are (in the static limit) not linear but parabolic in momentum while the linear bands remain unchanged and in particular ungapped.
Notice that contrary to the above case where $\theta_p = 45^\circ$
the positions of the dips in the average energies are nearly the same for both spin orientations.

For $\phi_k = 45^\circ$ (see the middle row of Fig.~\ref{FIG_spectrum_lin_pol}),
we observe remarkable gaps in all quasienergy spectra at $v_F k \approx 0.5\Omega$ and
$v_F k \approx \Omega$.
In addition, for finite SOIs an additional small gap opens up at the \textit{K} point separating
the valence and conduction bands; see Figs.~\ref{FIG_spectrum_lin_pol}(g) and \ref{FIG_spectrum_lin_pol}(h).
The time-averaged energies as shown in the middle row of Fig.~\ref{FIG_avg_energy_lin_pol}
resemble the circular result of Fig.~\ref{FIG_average_energy}.
The important differences, however, are the absence
[Figs.~\ref{FIG_avg_energy_lin_pol}(e) and \ref{FIG_avg_energy_lin_pol}(h)]
or reduction [Figs.~\ref{FIG_avg_energy_lin_pol}(f) and \ref{FIG_avg_energy_lin_pol}(g)] of the gap at the Dirac point and the
fact that the THz does not cause an additional spin splitting of the bands;
compare Fig.~\ref{FIG_spectrum}(b) and Fig.~\ref{FIG_spectrum_lin_pol}(f).
In contrast to the case of $\phi_k = 0^\circ$ the positions of the resonant dips
clearly split up for $\lambda_R \neq 0$.

Finally, for an in-plane angle perpendicular to the polarization direction,
i.e., $\phi_k = 90^\circ$, again in all four cases a distinct gap opens up at $v_F k \approx 0.5\Omega$.
While for $\lambda_R = 0$ the \textit{K} point energies do not change, 
a small gap opens up in the quasienergies in  Figs.~\ref{FIG_spectrum_lin_pol}(k)
and \ref{FIG_spectrum_lin_pol}(l) where $\lambda_R \neq 0$.
Furthermore, the dips in the mean energies of cases Figs.~\ref{FIG_avg_energy_lin_pol}(i) and \ref{FIG_avg_energy_lin_pol}(j)
are suppressed for $v_F k = \Omega$ but they are clearly present
in Figs.~\ref{FIG_avg_energy_lin_pol}(k) and \ref{FIG_avg_energy_lin_pol}(l).

\subsection{Density of states}
\begin{figure}[b]
\includegraphics[scale=0.32]{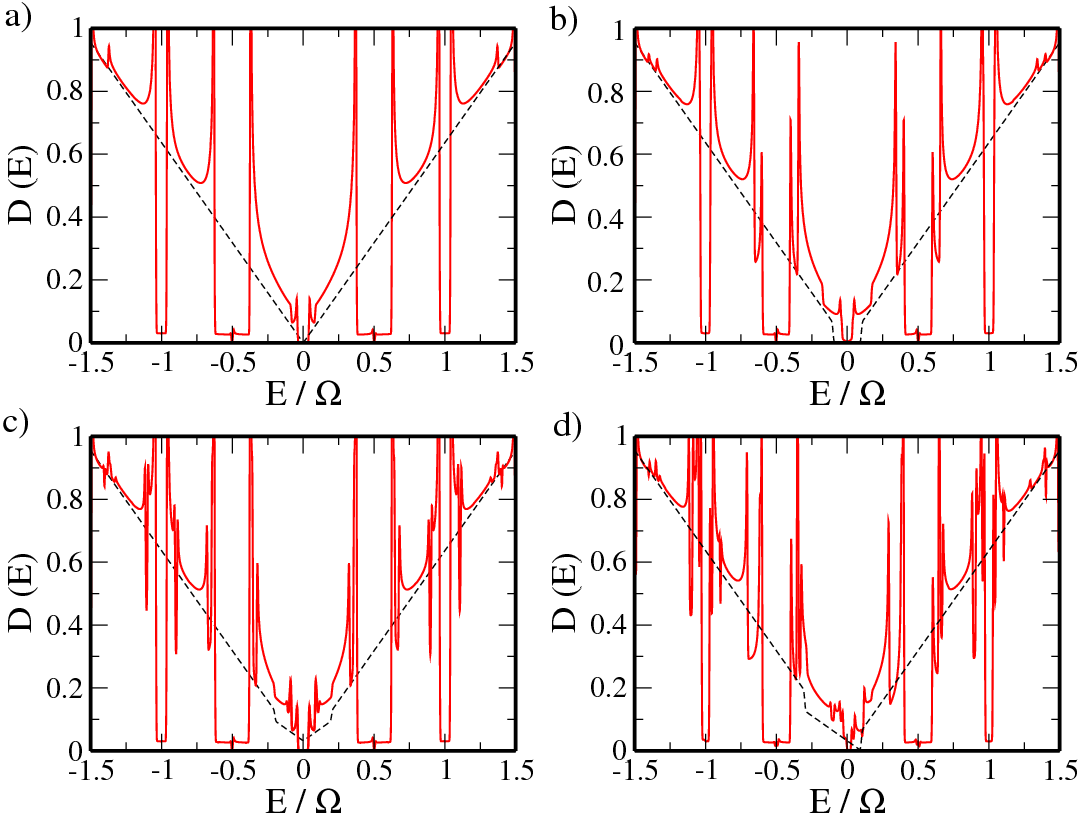}
\caption{(Color online) Time-averaged density of states calculated from Eq.(\ref{average_DOS})
under circularly polarized light ($\theta_p = 45^\circ$)
for various combinations of the SOC parameters:
$(\lambda_R/\Omega, \lambda_I/\Omega) = $ (a) $(0, 0)$, (b) $(0, 0.1)$, (c) $(0.1, 0)$, and (d) $(0.1, 0.1)$.
The field strength was set to $\alpha = 0.3$.
}
\label{FIG_DOS}
\end{figure}

\begin{figure}[t]
\includegraphics[scale=0.32]{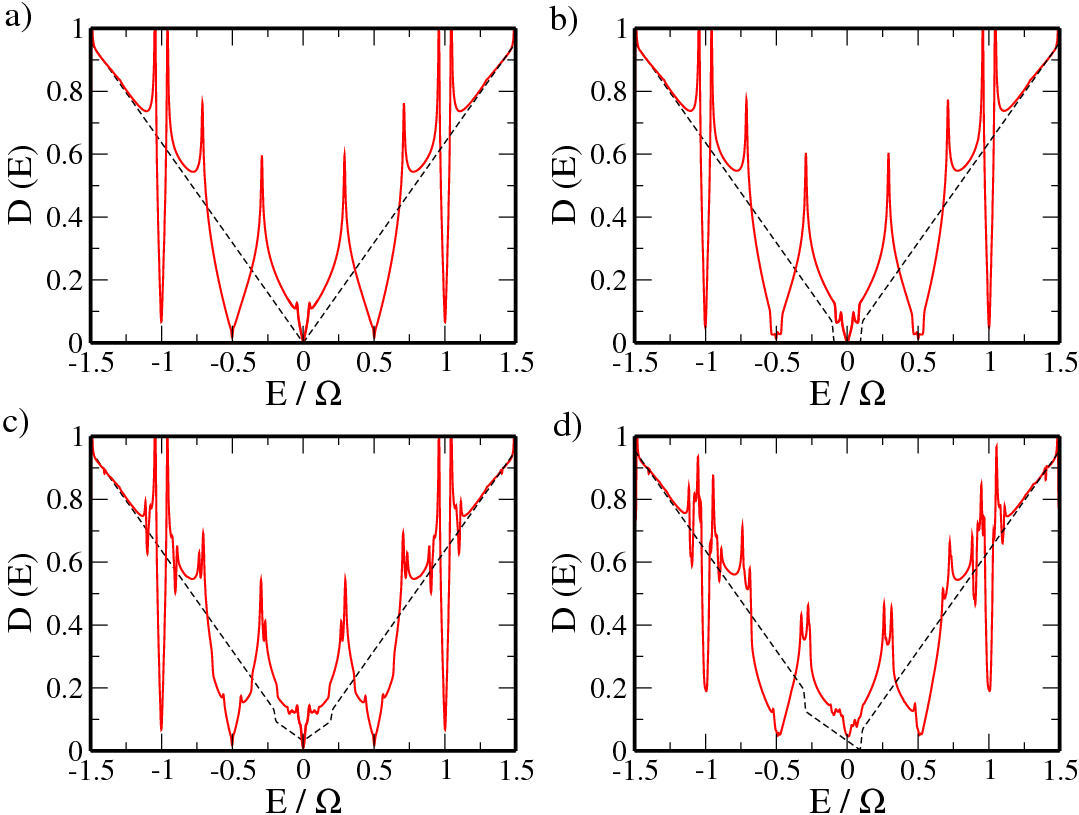}
\caption{(Color online) Time-averaged density of states calculated from Eq.~(\ref{average_DOS})
under linearly polarized light ($\theta_p = 0^\circ$)
for various combinations of the SOC parameters:
$(\lambda_R/\Omega, \lambda_I/\Omega) = $ (a) $(0, 0)$, (b) $(0, 0.1)$, (c) $(0.1, 0)$, and (d) $(0.1, 0.1)$.
The field strength was set to $\alpha = 0.3$.
}
\label{FIG_DOS_lin}
\end{figure}

In Figs.~\ref{FIG_DOS} and \ref{FIG_DOS_lin} the time-averaged DOS,\cite{Zhou_2011}
\begin{align}
D(E) &= g_v \sum_{\vec k, \mu\nu} \sum_{n=-\infty}^\infty
\left\langle \chi_{\vec k,\mu\nu}^n \vert \chi_{\vec k,\mu\nu}^n \right\rangle
\delta\left[E - \varepsilon_{\vec k,\mu\nu} + n\Omega \right] ,
\label{average_DOS}
\end{align}
is shown for various combinations of the SOC parameters
with (red solid line) and without (black dashed) electromagnetic field
for circularly and linearly polarized light, respectively.
The field amplitude was set to $\alpha = 0.3$.
The prefactor $g_v = 2$ is due to the valley degeneracy.

The static DOS for zero energy, shown as the dashed lines
in Figs.~\ref{FIG_DOS} and \ref{FIG_DOS_lin},
is zero in (a) and (b) and finite in (c) and (d).
In the latter case ($\lambda_R = \lambda_I$)
the charge neutrality point is shifted to $\lambda_{R/I}$.
The electromagnetic field yields a finite weight
$\left\langle \chi_{\vec k,\mu\nu}^n \vert \chi_{\vec k,\mu\nu}^n \right\rangle$
to the subbands in the first BZ even for momenta $v_F k > 0.5\Omega$.
This leads to a distinct increase of the DOS for small energies compared
to the field-free situation.\cite{Calvo_2011}
In Fig.~\ref{FIG_DOS_lin}(b), for example, the DOS
is greatly enhanced for $\abs E < \lambda_I$,
while in the static case $D(E) = 0$ in this regime.

As the quasienergies $\varepsilon_{\vec k,\mu\nu}$ have several extrema located
at $v_F k \approx \pm \Omega / 2$ and $\pm \Omega$ 
(see Figs.~\ref{FIG_spectrum} and \ref{FIG_spectrum_lin_pol}),
the DOS exhibits pronounced Van Hove singularities.\cite{Zhou_2011, Calvo_2011}
While due to the isotropy of the quasienergy spectrum in the case of circularly polarized light these singularities
occur for arbitrary angles of $\phi_k$,
for a linearly polarized field not all angles lead to Van Hove singularities.
As a consequence, the associated peaks rise much more strongly
for $\theta_p = 45^\circ$ compared to $\theta_p = 0^\circ$.
In the former, the DOS drops down almost vertically and remains roughly constant around $v_Fk \approx 0.5\Omega$ and $\Omega$.
This is in clear contrast to the linearly polarized case, where the decrease of the DOS 
is much smoother and the DOS becomes peaked, with $D(E)$ being roughly linear around
$v_Fk \approx 0.5\Omega$ and $\Omega$.\cite{Morell_2012}
Moreover, if spin degeneracy is lifted,
the DOS shows additional dips in between neighboring Van Hove singularities.
This is also true in Fig.~\ref{FIG_DOS}(b), where the splitting is caused by the THz field and not by the
Rashba term.

In the static limit signatures of SOIs in the DOS can be seen only
in a narrow region with $E \lesssim 0.25 \Omega $,
while for larger energies it is virtually the same in all cases
(see the dashed lines in Figs.~\ref{FIG_DOS} and \ref{FIG_DOS_lin}).
This changes once the field is switched on.
Here SOC manifests itself even for larger energies.
Comparing, e.g., Figs.~\ref{FIG_DOS}(a) and \ref{FIG_DOS}(c),
we see a remarkable difference even for energies $E \approx \Omega$
due to the additional dips and peaks in the DOS.
This can be understood from the quasienergy spectrum, e.g., in Fig.~\ref{FIG_spectrum}(c),
where due to the breaking of spin degeneracy several nearby points with a horizontal dispersion exist.
For circularly polarized light qualitatively the same happens also for the case of a purely intrinsic coupling ($\lambda_R = 0$)
as the bands split up for $\alpha \neq 0$.
However, this splitting is significant only for small momenta and hence
the multiple dips in Fig.~\ref{FIG_DOS}(b) can be seen only for energies
around $E \approx 0.5\Omega$.

\begin{figure}[t]
\includegraphics[scale=0.35]{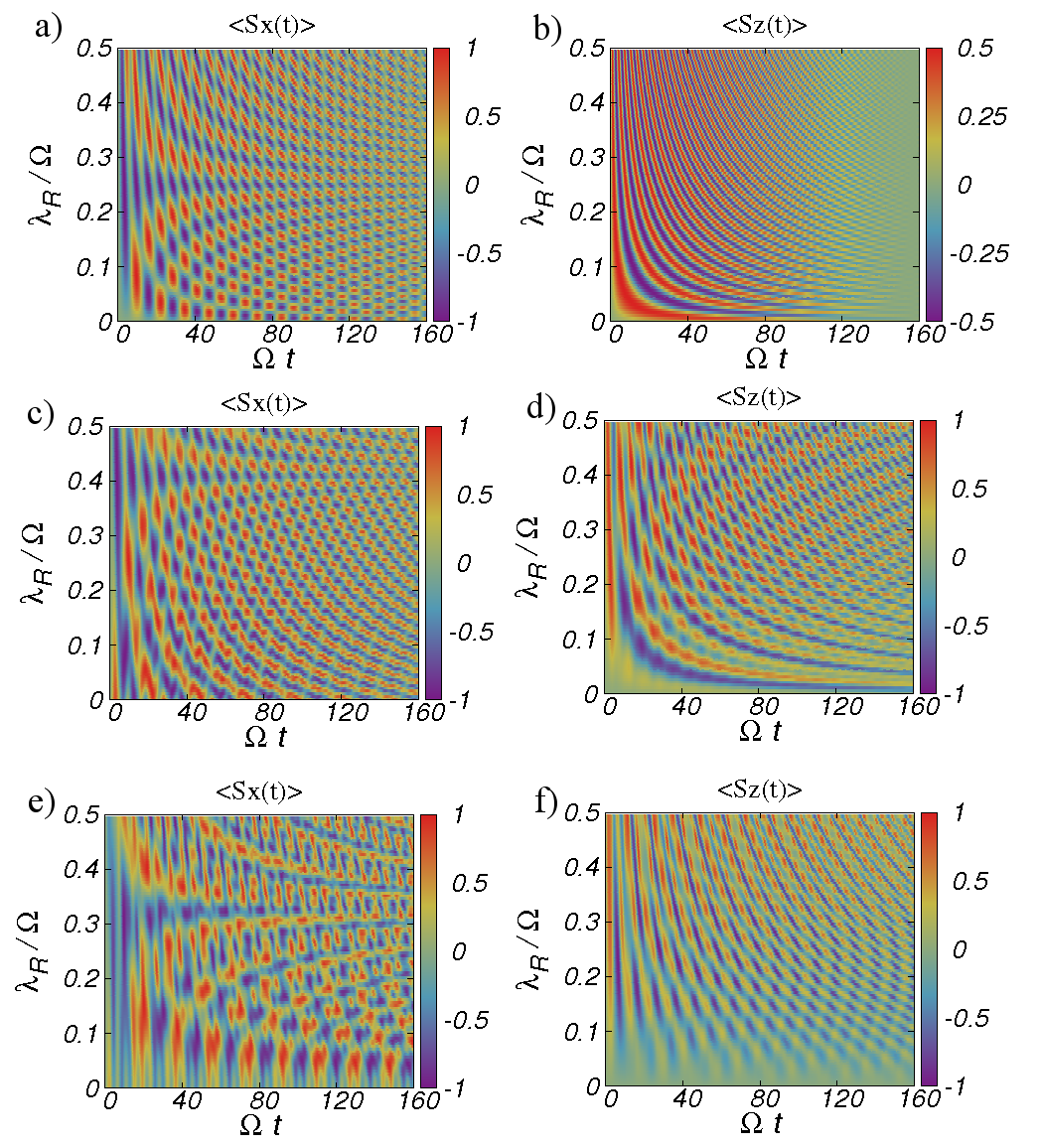}
\caption{(Color online) Time evolution of the $x$ and $z$ components of the spin polarization
without electric field (top row),
under a linearly polarized field ($\theta_p = 0^\circ$, $\alpha = 0.5$) (middle row),
and for circular polarization ($\theta_p = 45^\circ$, $\alpha = 0.5$) (bottom row),
as a function of the Rashba coefficient.
Parameters: $\lambda_I = 0.25\Omega$, $k = 0$.
}
\label{FIG_spin_pol_density_plot}
\end{figure}

\section{Spin polarization and wave packet dynamics}
\label{Sec:Dynamics}
We now discuss the dynamics of the real spin 
expressed by the operator $\hat S_{H,j}(t) = \sigma_{H, 0} s_{H, j}(t)$ ($j\in\{x,y,z\}$).
We restrict ourselves to an initial state described by a Gaussian wave packet
for a single momentum,\cite{Schliemann_2008} which is appropriate for a sufficiently broad
initial wave packet:\cite{Schliemann_2007}
\begin{align}
\left\langle \vec r \big\vert\Phi_{in}(t_0) \right\rangle = 
\frac 1{\sqrt\pi d} e^{-\frac{r^2}{2d^2}} 
\begin{pmatrix} \eta_1 \\ \eta_2 \\ \eta_3 \\ \eta_4 \end{pmatrix}
\label{initial_state} .
\end{align}
In the following the spinor components in Eq.~(\ref{initial_state}) are chosen as
$\eta_1 = -i\eta_2 = i\eta_3 = \eta_4 = 0.5$, i.e.,
the initial state is in general a linear combination of the static eigenvectors.
Because of 
\begin{align*}
\frac d{dt} \hat S_{H,z}(t) = -2\lambda_R \left[ \sigma_{H,x}s_{H,x}(t) + \sigma_{H,y} s_{H,y}(t) \right] ,
\end{align*}
changes in the initial out of plane spin polarization (SP)
\begin{align*}
\langle S_z(t_0) \rangle = \abs{\eta_1}^2 + \abs{\eta_2}^2 - \abs{\eta_3}^2 - \abs{\eta_4}^2 ,
\end{align*}
where $\langle . \rangle := \left\langle \Phi_{in} \vert . \vert \Phi_{in} \right\rangle$,
can be induced only if the Rashba contribution is finite.
Similarly, for the other two spin directions,
\begin{align*}
\langle S_x(t_0) \rangle = 2\Re{\bar\eta_1 \eta_3 + \bar\eta_2 \eta_4}
\end{align*}
and
\begin{align*}
\langle S_y(t_0) \rangle = 2\Im{\bar\eta_1 \eta_3 + \bar\eta_2 \eta_4} ,
\end{align*}
whose dynamics are described by
\begin{align*}
\frac d{dt} \hat S_{H, x/y}(t) = 2 \left[ \lambda_R \sigma_{H, x/y} s_{H,z}(t) \mp \lambda_I \sigma_{H, z} s_{H, y/x}(t) \right] ,
\end{align*}
at least one of the SOC coefficients has to be nonzero in order
to get a nontrivial time evolution.
In the following we set without loss of generality $t_0 = 0$.

\begin{figure}[b]
\includegraphics[scale=0.3]{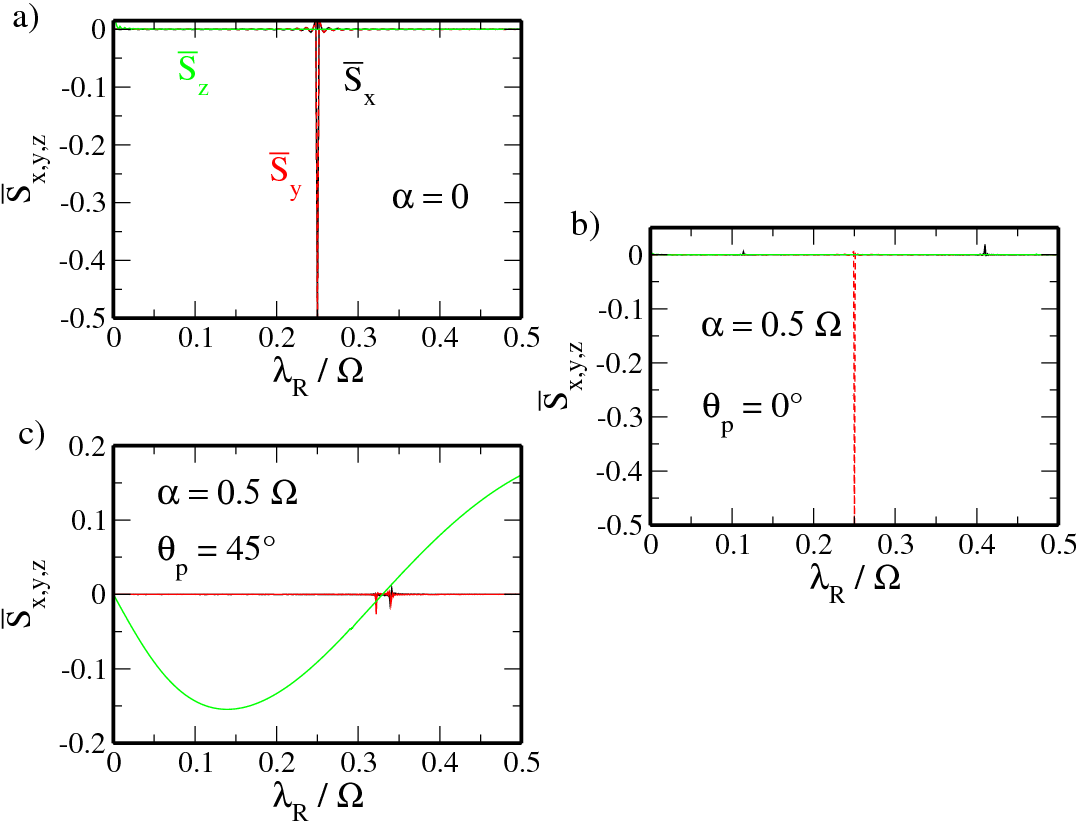}
\caption{(Color online) Mean spin polarization as a function of the Rashba parameter
(a) without electric field,
(b) under a linearly polarized field ($\theta_p = 0^\circ$, $\alpha = 0.5$),
and (c) for a circular polarization ($\theta_p = 45^\circ$, $\alpha = 0.5$).
The total simulation time is $\Omega t = 10 \, 000$.
Parameters: $\lambda_I = 0.25\Omega$, $k = 0$.
}
\label{FIG_spin_pol}
\end{figure}

\begin{figure}[t]
\includegraphics[scale=0.31]{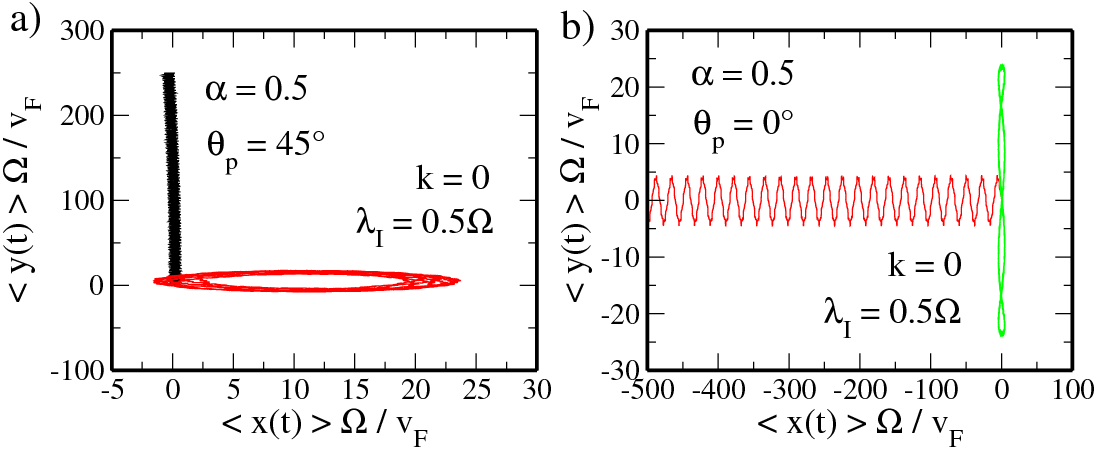}
\caption{(Color online) Orbital dynamics $\langle \vec r(t) \rangle$
calculated for a total simulation time of $\Omega t = 1000$
for (a) circularly ($\theta_p = 45^\circ$)
and (b) linearly ($\theta_p = 0^\circ$) polarized light 
for various Rashba SOC coefficients: 
$\lambda_R = 0$ (black line), $0.5$ (red), and $0.6$ (green).
Parameters: $\lambda_I = 0.5\Omega$, $\alpha = 0.5$, $k = 0$.
}
\label{FIG_Position}
\end{figure}

In Fig.~\ref{FIG_spin_pol_density_plot} we fix the intrinsic parameter $\lambda_I = 0.25\Omega$
and vary the Rashba constant at the Dirac point.
The field strength is set to $\alpha = 0.5$.
While for $\lambda_R \neq \lambda_I$ the in plane SP of the static system,
exemplarily shown for the $x$ component in Fig.~\ref{FIG_spin_pol_density_plot}(a),
shows fast oscillations around zero; right at the point $\lambda_R = \lambda_I$
the expectation values $\langle S_{x}(t) \rangle$ and $\langle S_{y}(t) \rangle$
oscillate around a finite value.
Subsequently the mean polarization $\bar S_{x/y}$ shown as black and red lines in Fig.~\ref{FIG_spin_pol}(a),
calculated for a total simulation time of $\Omega t = 10 \, 000$, vanishes or is very small for $\lambda_R \neq \lambda_I$,
while $\bar S_{x/y} = -0.5$ for $\lambda_R = \lambda_I$.
If we now turn on the THz field, the time evolution of the spin operators 
clearly becomes more complicated; see Figs.~\ref{FIG_spin_pol_density_plot}(c) and \ref{FIG_spin_pol_density_plot}(e).
If the field is linearly polarized along the $x$ direction, $\bar S_y$
is finite only for $\lambda_R = \lambda_I$, with $\bar S_y = -0.5$ at that point, as in the static case.
However, this is no longer true for $\bar S_x$ as can be seen from Fig.~\ref{FIG_spin_pol}(b)
where the peak for the $x$ component disappears.
For circularly polarized light both the peaks for $\bar S_x$ and $\bar S_y$ at $\lambda_R = \lambda_I$ vanish
and a significantly reduced dip at $\lambda_R \approx 0.34\Omega$ appears.
The out-of-plane SP ($\bar S_z$) of the static system oscillates around zero, where
the period of the oscillations increases for larger $\lambda_R$; see Fig.~\ref{FIG_spin_pol_density_plot}(b).
Hence in contrast to $\bar S_{x/y}$ the mean polarization $\bar S_z$ vanishes for arbitrary $\lambda_R$,
as can be seen from the green line in Fig.~\ref{FIG_spin_pol}(a).
This remains true for linearly polarized light where $\bar S_z \approx 0$ in all cases.
Compared to that the situation for circularly polarized light is quite different.
Here the $z$ component of the averaged spin oscillates as a function of $\lambda_R$
and, depending on the magnitude of the Rashba parameter, $\bar S_z$ can be either positive or negative
or zero for $\lambda_R = 0$ and $\lambda_R \approx 0.33\Omega$.
Notice that even though the intrinsic parameter has been fixed to $\lambda_I = 0.25\Omega$
in the above discussion, our findings remain qualitatively the same for other values of $\lambda_I$,
and in particular the peak, e.g., in Fig.~\ref{FIG_spin_pol}(a)
always appears right at the point where $\lambda_R = \lambda_I$.
A possible way to detect the SP has been described in Ref.~\cite{Rugal_2004}.
Here the sample is scanned by a cantilever in magnetic resonance force microscopy,
where the detected shift in frequency turns out to be related to the SP.

%%%%%%%%%%%%%%%%%%%%%%%%%%%%%%%%%%%%%%%%%%%%%%
%%%%%%%%%%	Position operator	%%%%%%
%%%%%%%%%%%%%%%%%%%%%%%%%%%%%%%%%%%%%%%%%%%%%%
The time evolution of the position operators in Heisenberg representation is given by
\begin{align}
\frac {d}{dt} \hat{\vec r}_H (t) =  i \left[ \hat H, \hat{\vec r}_H \right] = v_F \vec \sigma_H s_{H,0}(t) .
\label{Orb_dyn}
\end{align}
Note that contrary to electron and hole gas systems\cite{Schliemann_2005, Schliemann_2006, Schliemann_2007} the dissipative term
proportional to momentum is missing in Eq.~(\ref{Orb_dyn})
due to the Dirac-like nature of the charge carriers in graphene.
By calculating the usual velocity operator $\hat{\vec v}_H (t) = v_F \vec \sigma_H s_{H,0}(t)$
it is thus possible to extract the orbital dynamics of the system,
\begin{align*}
\langle \vec r(t) \rangle := \langle \Phi_{in} \vert \hat {\vec r}_H(t) \vert \Phi_{in} \rangle ,
\end{align*}
with respect to an initial wave packet given in Eq.~(\ref{initial_state}).
In Fig.~\ref{FIG_Position} this is shown for circularly (a) and linearly
(b) polarized light of strength $\alpha = 0.5$ for fixed $\lambda_I = 0.5\Omega$
and two different values of the Rashba SOC parameters for a total simulation time of $\Omega t = 1000$.
While for $\theta_p = 45^\circ$ and $\lambda_R \neq 0$ the trajectory resembles an ellipse,
and hence the particle becomes localized, as exemplarily
shown in the red curve in Fig.~\ref{FIG_Position}(a) for $\lambda_R = 0.5\Omega$,
the basic propagation is along the $y$ direction if the Rashba contribution vanishes and,
compared to $\langle y \rangle$, only moderate deviations
from the initial position in the $x$ direction can be seen.

For $\theta_p = 0^\circ$ and $\lambda_R = 0.5\Omega$ [see the red curve in 
Fig.~\ref{FIG_Position}(b)],
the main dynamics is along the $x$ axis with small oscillations
around $\langle y\rangle = 0$,
while in the other case of $\lambda_R = 0.6\Omega$ (green line) the trajectory is again bounded
in a finite region around $\langle x \rangle = \pm 5 v_F / \Omega$ and $\langle y \rangle = \pm 25 v_F/\Omega$,
respectively.

\section {Optical conductivity}
\label{Sec:Conductivity}

In this section the optical conductivity of irradiated graphene is calculated.
As we are not interested in processes that appear right after or before the
THz field is turned on and off, we consider the system in a quasi stationary
state and assume the probability distribution to be of the form
$P \propto e^{-\beta \bar\varepsilon_{\vec k,\mu\nu}}$,
where $\bar\varepsilon_{\vec k,\mu\nu}$ are the average energies introduced in Eq.~(\ref{Mean_energy})
and $\beta = 1 / T$ the inverse temperature.
The quasi equilibrium density matrix in the basis of the Floquet states then reads\cite{Zhou_2011, Gupta_2003, Faisal_1996, Hsu_2006}
$\left\langle \chi_{\vec k,\mu\nu} \vert \hat \rho^{qe} \vert \chi_{\vec k,\mu'\nu'} \right\rangle 
= \delta_{\mu,\mu'} \delta_{\nu,\nu'} f[\bar \varepsilon_{\mu,\nu}(\vec k)]$.
In the following, we restrict ourselves to zero temperature such that
the Fermi distribution function reads $f[E] = \heaviside{E_F - E}$,
with $E_F$ being the chemical potential.
The expression for the dissipative part of the time-averaged longitudinal optical conductivity, 
obtained from the nonequilibrium Green's function method derived in Ref.~\cite{Zhou_2011},
then reads
\begin{align}
& \Re{\bar\sigma_{xx}(\omega)} = \frac{g_v \pi e^2}{\omega} \sum_{\vec k, m, j} \sum_{\mu,\nu,\mu',\nu'}
\left\vert \left\langle \chi^{n-j}_{\vec k, \mu'\nu'} \Big\vert \hat v_x \Big\vert \chi^n_{\vec k,\mu\nu}
\right\rangle \right\vert^2 \notag \\
& \times \left( f\left[\bar\varepsilon_{\vec k,\mu\nu}\right] - f\left[\bar\varepsilon_{\vec k,\mu'\nu'}\right] \right)
\delta\left[ \omega + \varepsilon_{\vec k,\mu\nu} - \varepsilon_{\vec k,\mu'\nu'} - j\Omega\right] .
\label{Opt_cond}
\end{align}
The quasienergies and states entering Eq.~(\ref{Opt_cond}) are chosen to be in the first BZ,
although any other choice is possible as well.
From the $\delta$ function in Eq.~(\ref{Opt_cond}) we can see that, in principle,
transitions between all kinds of subbands are possible.
In the static limit only those subbands that correspond to the energies of Eq.~(\ref{Static_energies})
have a nonzero weight and Eq.~(\ref{Opt_cond}) reproduces previous results.\cite{Stauber_2008, Ingenhoven_2010, Scholz_2011}
For a finite driving the weight of the other subbands becomes nonzero,
whereas it increases for larger driving amplitudes,
and hence additional transitions become possible.

\begin{figure}[b]
\includegraphics[scale=0.3]{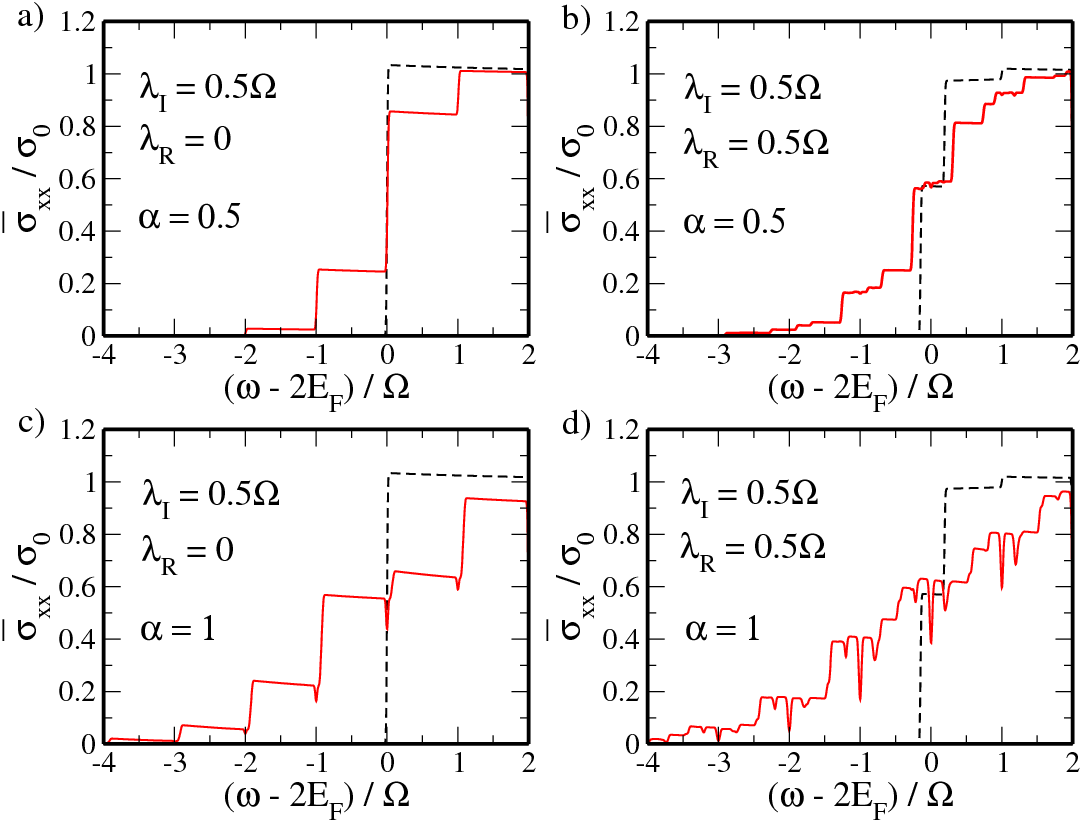}
\caption{(Color online) Optical conductivity under circularly polarized light ($\theta_p = 45^\circ$)
for various field strengths $\alpha = 0$ (black dashed curves), $0.5$ [red lines in (a) and (b)],
and $1.0$ [(c) and (d)],
and SOC parameters in units of $\sigma_0 = e^2 / 4$.
}
\label{FIG_opt_cond}
\end{figure}

\begin{figure}[t]
\includegraphics[scale=0.3]{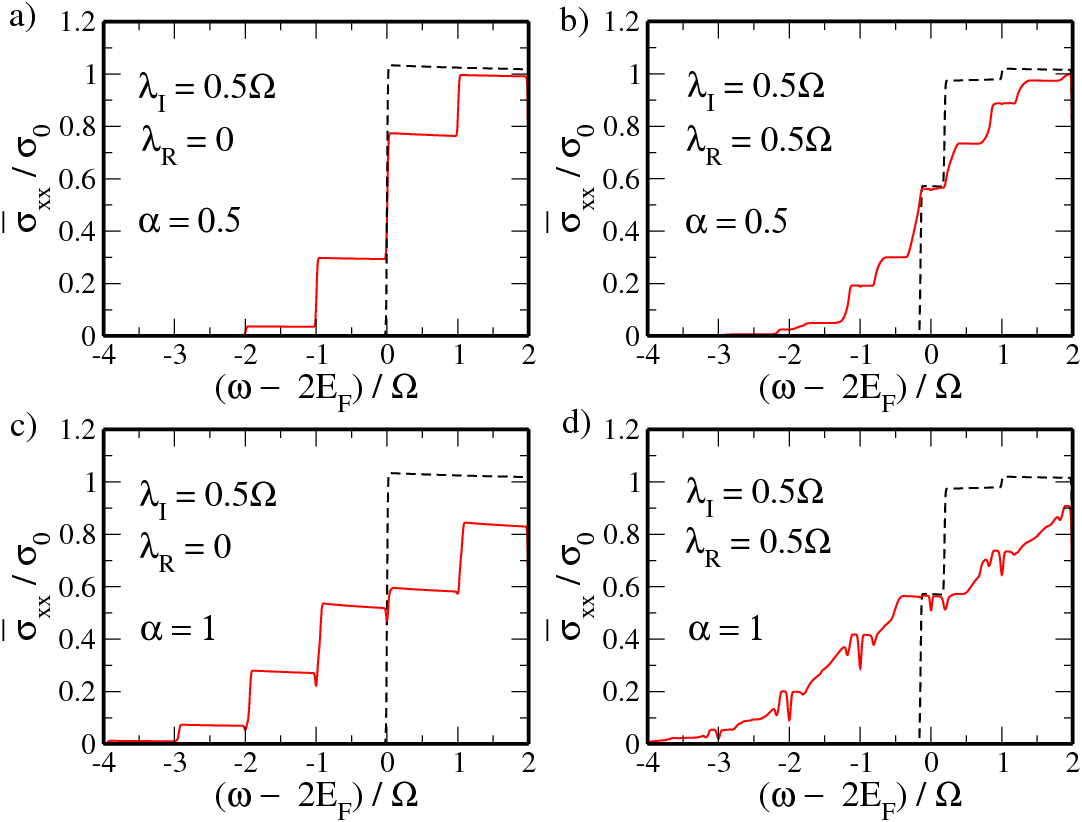}
\caption{(Color online) Optical conductivity under linearly polarized light ($\theta_p = 0^\circ$)
for various field strengths $\alpha = 0$ (black dashed curves), $0.5$ [red lines in (a) and (b)],
and $1.0$ [(c) and (d)],
and SOC parameters in units of $\sigma_0 = e^2 / 4$.
}
\label{FIG_opt_cond_lin_pol}
\end{figure}

In Figs.~\ref{FIG_opt_cond} and \ref{FIG_opt_cond_lin_pol} we show the optical conductivity
calculated for a fixed Fermi energy of $E_F = 3\Omega$ under the influence of
circularly and linearly polarized light, respectively.
The field strength is $\alpha = 0$, $0.5$, and $1.0$.
The main feature of the static conductivity,
as shown, e.g., in the dashed curve in Fig.~\ref{FIG_opt_cond}(a),
is its steplike behavior at $\omega = 2E_F$,
where transitions from the valence to the conduction band become possible.
Switching on the time-dependent field leads to 
several additional steps in $\bar\sigma_{xx}$,\cite{Zhou_2011}
due to photon-assisted processes.
By comparing e.g., Figs.~\ref{FIG_opt_cond}(a) and \ref{FIG_opt_cond}(c), it becomes clear
that the number of steps increases for larger coupling strengths $\alpha$
as the weight is distributed over a broader range of subbands.
The effect of the Rashba term, which leads to a distinct breaking of the spin degeneracy
of each subband, furthermore induces several intermediate steps
as the number of possible transitions in the $\delta$ function of Eq.~(\ref{Opt_cond}) becomes much larger.
From Figs.~\ref{FIG_opt_cond}(b) and \ref{FIG_opt_cond_lin_pol}(b)
we can see that the basic structure of $\bar \sigma_{xx}$ is the same for $\theta_p = 45^\circ$
and $\theta_p = 0^\circ$, but in the latter the conductivity turns out to be slightly smoother.
By increasing the field strength to $\alpha = 1.0$ we observe
dips in the conductivity at $\omega = n\Omega$,\cite{Zhou_2011}
where the effect is clearly larger for $\theta_p = 45^\circ$ than $\theta_p = 0^\circ$;
see, e.g., Figs.~\ref{FIG_opt_cond}(c) and \ref{FIG_opt_cond_lin_pol}(c).
The inclusion of the Rashba term creates further dips for slightly smaller and 
larger frequencies, respectively.
These dips are due to the appearance of gaps in the quasienergy spectrum
(see the discussion in Sec.~\ref{Sec:Spectrum}),
as some transitions are no longer possible.
From Figs.~\ref{FIG_opt_cond}(c) and \ref{FIG_opt_cond}(d) one can see that while
the static conductivities (dashed curves) are quite similar in both cases,
i.e., the effect of $\lambda_R$ is only slight,
remarkable differences occur in the driven case,
and hence SOC effects are greatly enhanced.

\section{Conclusions}
\label{Sec:Conclusions}

In this work the effect of a time-dependent electric field on a monolayer of graphene
including SOIs of the intrinsic and Rashba types has been studied.

We have demonstrated that a circularly polarized THz field can be used not only to induce a gap at the Dirac
point, which transforms graphene from a semimetal to an insulator,
but also to close an existing gap in the quasienergies.
In the opposite case of a linear polarization the spectrum turned out to be highly anisotropic
and, depending on the strength of the SOC parameters and on the orientation of the field,
gaps in the spectrum might appear at the \textit{K} point and at the photon resonances 
$v_F k \approx 0.5 n \Omega$, or become suppressed.

While the effect of SOIs on the DOS of the static sample could be seen only for 
energies $E \lesssim 0.25\Omega$, due to the existence of a multiple number of dips,
signatures of SOC in the DOS of irradiated graphene
appear even at much larger energies.

By introducing a time-dependent field it turned out to be possible to induce a finite
net spin polarization in the sample.
The sign and magnitude, e.g., of the out-of-plane polarization,
can be modulated by changing the ratio of the SOC parameters,
which can be done experimentally by adjusting the Rashba coefficient via an electric gate.

In the last part of this work the longitudinal optical conductivity was calculated.
As reported already in Ref.~\cite{Zhou_2011}, the conductivity of irradiated graphene
exhibits a multi step structure as transitions between a variety of subbands become possible.
The number of steps depends not only on the coupling strength, but also on the magnitude of
the Rashba parameter and on the polarization direction.
Furthermore, for large enough coupling strengths the conductivity drops down for frequencies
around the photon energy.\cite{Zhou_2011}
As for the DOS, compared to the static result the effect of SOIs on the optical conductivity
is greatly enhanced for $\alpha \neq 0$, which is mainly caused by the Rashba contribution.

Finally, let us point out that even though the SOC parameters within this work have been
chosen to be smaller than (but comparable to) the energy of the field,
our findings [such as the appearance of gaps in the quasienergy spectrum or the oscillatory
behavior of the out-of-plane spin polarization in Fig.~\ref{FIG_spin_pol}(c)]
are not limited to this case, but can also be observed in the opposite case of $\lambda_{R/I} \gtrsim \Omega$.

\acknowledgments
We thank M. Busl, M. Grifoni, S. Kohler, and M.~W. Wu, for useful discussions.
This work was supported by Deutsche Forschungsgemeinschaft via Grant No.~GRK 1570.

%%%%%%%%%%%%%%%%%%%%%%%%%%%%%%%%%%
%% 	bibliography		%%
%%%%%%%%%%%%%%%%%%%%%%%%%%%%%%%%%%

\end{document}